\documentclass[a4paper,fleqn,usenatbib]{mnras}
\usepackage{amsmath,amssymb}

\usepackage{upgreek}
\usepackage{txfonts}

\usepackage[T1]{fontenc}
\usepackage{ae,aecompl}

\usepackage{graphicx}	
\usepackage{amsmath}	
\usepackage{amssymb}	

\title[Magnetic support in molecular medium]{Quantifying the
interplay between gravity and magnetic field in molecular clouds --a  possible
multi-scale energy equipartition in NGC6334}

\author[Guang-Xing Li,  Andreas Burkert]{Guang-Xing Li$^{1}$\thanks{Contact e-mail:
gxli@usm.lmu.de} ,  Andreas
Burkert$^{1, 2}$
\\
$^{1}$University Observatory Munich, Scheinerstrasse 1, D-81679, M\"unchen,
Germany\\
$^{2}$ Max-Planck-Fellow, Max-Planck-Institute for Extraterrestrial
Physics, Giessenbachstrasse 1, 85758, Garching, Germany
}
\date{\today}
\pubyear{2015}

\begin{document}
\label{firstpage}
\pagerange{\pageref{firstpage}--\pageref{lastpage}}
\maketitle

\begin{abstract}
 The interplay between gravity, turbulence and the magnetic field determines the
 evolution of the molecular ISM and the formation of the stars. In spite of
 growing interests, there remains a lack of understanding of the importance of
 magnetic field over multiple scales. We derive the magnetic energy spectrum --
 a measure that constraints the multi-scale distribution of the magnetic energy,
 and compare it with the gravitational energy spectrum derived in Li \& Burkert
 (2016). In our formalism, the gravitational
 energy spectrum is purely determined by the surface density PDF, and the
 magnetic energy spectrum is determined by both the surface density PDF and the
 magnetic-field-density relation.  If regions
 have density PDFs close to $P(\Sigma)\sim \Sigma^{-2}$ and a universal magnetic
 field-density relation $B\sim \rho^{1/2}$, we expect a
 multi-scale near equipartition between gravity and the magnetic fields. This
 equipartition is found to be true in NGC6334 where estimates of magnetic
 fields over multiple scales (from 0.1 pc to a few parsec) are available.
 However, the current observations are still limited in sample
 size.
 In the future,
 it is necessary to obtain multi-scale measurements of magnetic fields from
 different clouds with different surface density PDFs and apply our formalism to further study the gravity-magnetic field interplay.

 \end{abstract}


\section{Introduction}


Star formation process is believed to be determined by a combination of
turbulence, gravity and the magnetic field.
A growing number of observations seem to indicate that the magnetic field can be
dynamically important \citep[e.g.][and references therein]{2014prpl.conf..101L}.
In previous studies, the importance of the magnetic field has been studied by
direct estimates of the the field strength
\citep{1999ApJ...520..706C,2015Natur.520..518L}, or by studying the correlation between field orientations and properties of the gas condensations
\citep{2008A&A...486L..13A,2015ApJ...799...74P,
2011MNRAS.411.2067L,2012MNRAS.420.1562H,2014ApJ...792..116Z,2016MNRAS.462.1517P,2016A&A...586A.136P,2017arXiv170203853S,2017ApJ...842L...9H}
\footnote{The results of \citet{2016MNRAS.462.1517P} have been
recently corrected in an erratum   \citep{2016MNRAS.462.2011P}.}, as well as by
studying the properties of the fragments \citep{2015A&A...584A..67B,2016A&A...593L..14F}.
However, since star formation  is a multi-scale process, it is necessary to
constrain the importance of the magnetic field over multiple scales.
Previously, it has been proposed by \citet{1988ApJ...326L..27M} that the
magnetic energy density is comparable to the energy density of the gravitational
field over multiple scales. Thanks to
many recent observational constraints as well as theoretical developments, one
can now expect to understand the interplay between gravity and magnetic field to a better
accuracy.

 In previous papers \citep{2016MNRAS.461.3027L,2017MNRAS.464.4096L}, we proposed
 to study the importance of gravity using a measure called the
 gravitational energy spectrum, which quantifies the multi-scale distribution of
 the gravitational energy of a molecular cloud.  The gravitational energy
 spectrum is mainly determined by the surface density structure of a cloud
 measured in terms of the surface density PDF. We found that the
 measured gravitational energy spectrum and the expected kinetic energy spectrum
 from turbulence cascade exhibit a multi-scale near-equipartition \citep{2016MNRAS.459.2432V,2016MNRAS.461.3027L,2017MNRAS.464.4096L}. However,
 the importance of the magnetic fields remains unconstrained. In this paper, we
 develop a new measure called the magnetic energy spectrum, where we constrain
 the distribution of magnetic field energy density over multiple scales by
 combining the surface density PDF and the magnetic field-density relation. The
 formalism allows us to study the interplay between gravity and the magnetic
 field in molecular clouds over multiple scales. We apply the method to the
 existing measurements and discuss observational perspectives.

\section{The formalism}\label{sec:formalism}
The general idea is to represent the magnetic energy
content of a molecular cloud in the $k$ space where $k\sim 1/l$ is the spatial
frequency. This allows us to study the importance of magnetic field over
multiple scales in different regions, and compare multi-scale magnetic field
energy density with the gravitational energy density. To achieve this, we
approximate a region being spherically symmetric. Following \citet{2016MNRAS.461.3027L},
we construct the effective radial density profile based on
the surface density PDF, and derive the gravitational energy spectrum, which
represents the multi-scale distribution of gravitational energy in a region. How
we achieve this will be briefly explain in Sec. \ref{sec:eg}. In Sec.
\ref{sec:eb}, by incorporating the magnetic field-density relation into our
model, we derive the magnetic energy spectrum, and in the next section we
compare gravitational and magnetic energy spectrum.
A list of definitions can be found in Table
\ref{tbl:def}.

\begin{table*}
\begin{center}
\begin{tabular}{ ccc }
 \hline
 Symbol & Name  & Definition  \\
\hline
$\Sigma$ & surface density &  \ldots \\
$\rho$ & density & \ldots \\
$B$ & magnetic field & \ldots \\
r & effective radius, scale & \ldots\\
k & wavenumber  &  $k\approx 1/r$\\
$N$ & number of regions & \ldots\\
$P(\Sigma)$ & surface density PDF  & \ldots\\

$E_{\rm  p}(k)$ & gravitational energy spectrum & Eq. \ref{eq:ep}\\
$E_{\rm p, corr}(k)$ & gravitational energy spectrum, corrected with the number
of regions& Eq. \ref{eq:ep:corr}\\
 $E_{\rm  B}(k)$ & magnetic energy spectrum & Eq. \ref{eq:eb}\\
$\rho(r)$ & effective radial profile & \ldots \\

 $\gamma_{\Sigma}$& slope of the surface density PDF & $P(\Sigma) \sim
 \Sigma^{- \gamma_{\Sigma}}$ \\

  $\gamma_{\rho}$ & slope of the effective radial profile & $\rho(r)
  \sim r^{- \gamma_{\rho}}$\\
  $\gamma_{\rm B}$  & slope of the magnetic field-density relation&
$B \sim \rho^{\gamma_{\rm B}} $\\
  $\Phi$ & magnetic flux & \ldots\\
  $\eta$  & mass-to-flux ratio & $\eta = M / \Phi$ \\
$\epsilon_{\rm p}$ & slope of the gravitational energy spectrum &
 $E_{\rm  p}(k) \sim k^{-\epsilon_{\rm p}}$ \\
  $\epsilon_{\rm B}$ & slope of the magnetic energy spectrum &
 $E_{\rm  B}(k) \sim k^{-\epsilon_{\rm B}}$\\
  \hline

\end{tabular}
\end{center}
\caption{\label{tbl:def} List of definitions of mathematical symbols.}
\end{table*}

\subsection{Gravitational energy spectrum}\label{sec:eg}
We
present a shortened version of the derivation of the gravitational energy
spectrum presented in \citet{2016MNRAS.461.3027L,2017MNRAS.464.4096L}.
For a region where the surface density PDF takes the form (where $\Sigma$
stands for the surface density)
\begin{equation}
P(\Sigma) \sim \Sigma^{- \gamma_{\Sigma}}\;,
\end{equation}
one can approximate it as spherically symmetric
and derive the effective radial profile of that region
\citep{2007ApJ...665..416K,2010A&A...512A..81F,2014ApJ...781...91G,2016MNRAS.461.3027L}
\begin{equation}\label{eq:rhor}
\rho(r) \sim  r^{- \gamma_{\rho}} \sim r^{- (1 + \frac{2}{\gamma_{\Sigma}})}\;,
\end{equation}
where $\rho$ is the density and $r$ is the radius.
One can compute the gravitational energy contained within an effective radius
$r$ \citep{2016MNRAS.461.3027L},
\begin{equation}
E_{\rm p} \sim G \; \frac{m^2}{ r} \sim G \rho^2 r^5 \sim r^{3 - \frac{4}
{\gamma_{\Sigma}}}\;.
\end{equation}
Substituting in $k \sim 1/r$ where $k$ is the wavenumber, we derive the
\emph{gravitational energy spectrum}

\begin{equation}\label{eq:ep}
E_{\rm p}(k) \sim \frac{\partial{E_{\rm p}}}{\partial r} \;
\frac{\partial{r}}{\partial k} \sim k^{\frac{4}{\gamma_{\Sigma}}-4},
\end{equation}
which is a representation of the gravitational energy of a region as a function
of the wavenumber $k$.
Molecular clouds typically have surface density PDFs similar to $P(\Sigma)
 \sim \Sigma^{-2}$ \citep[e.g. ][]{2015A&A...576L...1L}, from which we derive a fiducial
gravitational energy spectrum $E_{\rm p}\sim k^{-2}$.

\subsection{Magnetic energy spectrum}\label{sec:eb}
Based on observations \citep[e.g.][]{1999ApJ...520..706C},
we assume that beyond a critical density,  that the magnetic field
strength in a region can be described as a function of gas density, e.g. $B\sim
\rho^{\gamma_{\rm B}} $, where $B$ is the magnetic field and $\rho$ is the gas
density. The fiducial value of $\gamma_{\rm B} $ is around
0.5.
Combined with the results presented in Sec. \ref{sec:eg}, we derive the
field strength as a function of the effective radius:
\begin{equation}\label{eq:brho}
B \sim \rho^{\gamma_{\rm B}} \sim r^{- (1 + \frac{2}{\gamma_{\Sigma}})
\gamma_{\rm B}}\;,
\end{equation}
where we have used Eq. \ref{eq:rhor}. We can express the magnetic energy
enclosed within a region of radius $r$ as a function of the radius:
\begin{equation}\label{eq:br}
E_{\rm B} \sim B^2\, r^3 \sim \rho^{2 \gamma_{B}} r^3 \sim r^{3 - 2
\gamma_{\rho} \gamma_{\rm B}}\;.
\end{equation}
This allows us to
derive the magnetic energy density as a function of wavenumber $k$
(where $k \sim 1/r$)
\begin{equation}\label{eq:eb}
E_{\rm B}(k)\sim \frac{\partial{E_{\rm B}}}{\partial r} \;
\frac{\partial{r}}{\partial k}\sim k^{2 \gamma_{\rm B} \gamma_{ \rho}- 4}
\sim k^{2 \gamma_{\rm B} (1 + \frac{2}{\gamma_{\Sigma}})- 4}\;.
\end{equation}
$E_{\rm B}(k)$ is a representation of the distribution of magnetic energy
over multiple scales, and we call it the \emph{magnetic energy spectrum}.
Using fiducial values, $\gamma_{\Sigma} \approx 2$
\citep{2009A&A...508L..35K,2015A&A...576L...1L}, $\gamma_{\rm B} \approx 0.5$
\citep{1999ApJ...520..706C}, thus $E_{\rm B} \sim k^{-2}$.

%

\subsection{Accuracy of the close-to-spherical assumption}
Star-forming regions exhibit complicated geometries, and the
close-to-spherical geometry we adapt to derive the effective radial profile
and the subsequent measures might lead to some inaccuracies. However these
uncertainties are of the order of unity and are not significant. For example,
 the first effect that one must consider is the effect of aspect ratio. It
turns out that the amount of gravitational energy contained in a region is not sensitive to
 the assumed aspect ratio. This has been proven by
 \citet{1992ApJ...395..140B}.
 On the other hand, how the amount of the magnetic energy depends on the aspect
 ratio is related to the way the size of a region is defined.
 Consider a region where the three axes are ($r_{\rm a}, r_{\rm b}, r_{\rm c}$), the magnetic energy should be $B^2
 r_{\rm a} r_{\rm b} r_{\rm c}$ and the estimated magnetic energy should be $B^2
 r_{\rm mean}^3$. Whether Eq. \ref{eq:br} provides a good estimate to
 the magnetic energy depends on the difference between  $r_{\rm mean}^3$ and
 $r_{\rm a} r_{\rm b} r_{\rm c}$. This is related to how $r_{\rm
 mean}$ is defined:
 when the projected axes are $r_{\rm min}$ and $r_{\rm max}$, our formula is more accurate when one defines $r_{\rm mean}\approx \sqrt{r_{\rm min}
 r_{\rm max}}$. This should be taken into account in future reconstructions of
 the magnetic-field density relation.

The remaining uncertainty arises from  the existence of
subregions. This issue has been already discussed in
our previous papers: \citet{2016MNRAS.461.3027L} considered a thought experiment
where one artificially splits an object of mass $m$ into several subregions and keeps the density unchanged.
Then they estimate the change of the gravitational energy as the result of this artificial
fragmentation process. Through this they can access the effect of substructures
on the estimated energy. When the region has been split into $N$
identical subregions of the same density, the total gravitational energy should
scale as $N^{-2/3}$ where $N$ is the number of regions on a given scale \footnote{One can easily understand this scaling:
the mass of a subregions scales with $N^{-1}$ and the size of the regions scales
with $N^{-1/3}$. The total gravitational energy of one region scales with
$N m^2/r \sim N^{-2/3}$.}. In general, splitting a big region into smaller
regions decreases the total gravitational energy.
Observationally, one can measure the number of subregions as a function of the scale $N = N(k)$ where $k\approx 1 / l$ is the wavenumber, $l$ is the scale, and the gravitational energy spectrum of the
system is \citep{2016MNRAS.461.3027L}
\begin{equation}\label{eq:ep:corr}
E_{\rm p, corr}(k) = k^{\frac{4}{\gamma_{\Sigma}}-4} N(k)^{-2/3}\;,
\end{equation}
where $E_{\rm p, corr}$ stands for the corrected gravitational energy spectrum
where the number of subregions as a function of the wavenumber $k$ is taken into
account as a correction term.
In our previous papers \citep{2016MNRAS.461.3027L,2017MNRAS.464.4096L}, we
found that assuming $N=1$ already gives results that are reasonably accurate
(where slope differs by roughly 5\%).
When the number of subregions is well-defined, one can always apply this
correction to improve the accuracy of the analysis.

Under the assumption that the magnetic field strength is a monotonic function of
the gas density, one can prove that the magnetic energy does not
scale with the number of subregions and thus needs no additional correction. To
demonstrate this, we consider a similar thought experiment where one region has
been split into $N$ subregions of the same density. We also assume the
density-magnetic field relation (Eq. \ref{eq:brho}). Before the artificial
fragmentation, the magnetic energy is $E_{B}\approx B^2 r^3 $. After the
artificial fragmentation, the subregions have masses $m'$ and radii $r'$.
We have $r' = r  N^{-1/3}$, $m'=m / N$. Using
the magnetic field-density relation, the total
magnetic energy $E_{\rm B}' = N B'^2 r'^3  = B^2 r^3$ is unchanged. Thus
the formula for the magnetic energy spectrum is accurate even if the region has
fragmented.
Since in our formalism, the magnetic energy spectrum is derived in a way
that is similar to the derivation of the gravitational energy spectrum presented
in \citet{2016MNRAS.461.3027L,2017MNRAS.464.4096L}, in general, we expect the
uncertainty of the magnetic energy spectrum slope to be similar to the
uncertainty of the gravitational energy spectrum. For observations that cover
two orders of magnitudes in scale, we expect an uncertainly 5 \%
in the estimated slope of the magnetic energy spectrum due to our
reconstruction.

Recently, the Herschel satellite revealed the ubiquitous existence of filamentary
structures in molecular clouds \citep[e.g.][and
references therein]{2014prpl.conf...27A}.
However, we do not expect this to have a significant impact on our results. In
essence, the filaments are simply crests of the underlying density
distributions, and in many cases (depending on how one defines them) they represent small-scale structures that superimposed on a large-scale density gradient, and our energy terms are largely
contributed from the large-scale density gradient. Therefore, we can neglect
these filamentary structures for our purpose. We note that this simplification seems to be
justified also by the fact that our analytical formula for the gravitational
energy energy spectrum can already provide a good description to the gravitational
energy spectrum constructed directly from the observations
\citep{2017MNRAS.464.4096L}.

\subsection{Equipartition condition}\label{sec:equi}
A special case to consider is the exact equipartition
between gravitational and magnetic energy, where
\begin{equation}
E_{\rm B} \sim E_{\rm p}\;.
\end{equation}
 Using  Eqs. \ref{eq:rhor} and \ref{eq:br} and neglect the dependence of $N$ on
 $r$ , we have
\begin{equation}\label{eq:equi}
\gamma_{\rm B} = \frac{2}{\gamma_{\Sigma} + 2}\;.
\end{equation}
This is the approximate condition for the multi-scale equipartition between
magnetic and gravitational energy, where the slope of the
magnetic field-density relation is connected to the power-law slope of the
surface density PDF.

\subsection{Connection to the mass-to-flux ratio}\label{sec:flux}
A often-used diagnostics of magnetic field strength is the mass-to-flux ratio.
In our model, assuming that the field lines are approximately straight, we
 estimate the mass-to-flux ratio as a function radius $r$.
Combining Eq. \ref{eq:rhor}  and Eq. \ref{eq:brho},
\begin{equation}\label{eq:eta}
\eta = \frac{M}{\Phi} \sim \frac{\rho r^3}{B r^2} \sim r^{1 -
\gamma_{\rho} + \gamma_{\rho}\gamma_{\rm B}} = r^{ 2(\gamma_{\rm B} - 1
)/\gamma_{\Sigma} + \gamma_{\rm B}}\;,
\end{equation}
where $\Phi$ is the magnetic flux and $\eta$ is the mass-to-flux ratio.
A special case to consider is when the magnetic energy and gravitational energy
reach exact equipartition. Substituting in Eq. \ref{eq:equi}, one finds that the
mass-to-flux ratio is independent on the radius ($\eta \sim r^0$). Thus, when the field lines
are not tangled, the equipartition condition discussed in Sec \ref{sec:equi}
implies a constant mass-to-flux ratio that does not evolve with the scale.

\section{Magnetic support in observations}
We apply our formalism to observations and study the effectiveness of magnetic
support. This is achieve by studying the relation
between the gravitational energy spectrum and the magnetic energy spectrum.


In the state-of-the-art observations, even though the slopes of the
surface density PDFs are relatively well-constrained, observational
constraints on the slopes of the magnetic field-density relations are
still limited. In
\citet{1999ApJ...520..706C}, the authors obtained 27 direct Zeeman measurements from different regions. The original analysis of
\citet{1999ApJ...520..706C}
yield a magnetic field-density relation of $B\sim \rho^{0.47\approx 1/2}$.
Later, the authors
\citep{2010ApJ...725..466C} refined the analysis using a fully Bayesian
approach, and found $B\sim\rho^{0.65\pm 0.05}$ for  $n_{\rm H_2}>300\; \rm cm^{-3}$. Although the
more recent results of \citet{2010ApJ...725..466C}  is generally considered as
being more reliable
 compared the older results of \citet{1999ApJ...520..706C}, both neglects the
 dependence of the magnetic field density relation on the density PDF, which,
 according to our analysis, is crucial. As a result, it is difficult to judge
 whether $B\sim\rho^{0.65\pm 0.05}$  is superior compared to $B\sim
 \rho^{0.47\approx 1/2}$ or not. Recently, \citet{2015MNRAS.451.4384T}
 revisited the observational biases and argued that $B\sim \rho^{1/2}$ is still the preferred
relation. However, the analysis also shares the drawback that they did not
consider the interplay of the magnetic field density relation with the density
PDFs. Realising these uncertainties, in Sec. \ref{sec:obs:general}, we discuss
the effectiveness of magnetic supports under different magnetic field-density relations.

According to our formalism, to constrain the effectiveness of magnetic support
over multiple scales, one necessarily needs to obtain multi-scale measurements
of magnetic field strengths in individual regions and combine the magnetic
field-density relation with the density PDF measurements.
However, such observations have been limited. To our knowledge, measurements on
NGC6334 preformed by \citet{2015Natur.520..518L}  is the only case where the
magnetic field over multiple scales are estimated.
The authors found that $B\sim \rho^{0.4\pm 0.04}$. However, one should note that
these are not direct measurements of field strengths: the magnetic field
strength is estimated by combining preexisting measurement of field strength on
10 pc scale with dust polarisation measurements that constraints the field
orientations. The authors used the force balance arguments recursively to derive
the field strength on smaller scales. For the estimates to be valid, the system
must be close to virial equilibrium.
Therefore, the results are model-dependent. In Sec.
\ref{sec:ngc6334}, we provide a detailed study of the effectiveness magnetic
support in NGC6334. A list of the derived slopes can be found in
Table. \ref{tbl:results}.

\subsection{General importance of magnetic support under $B\sim \rho^{1/2}$ and
$B\sim \rho^{0.65}$ }
\label{sec:obs:general}
\begin{figure*}
\includegraphics[width=1 \textwidth]{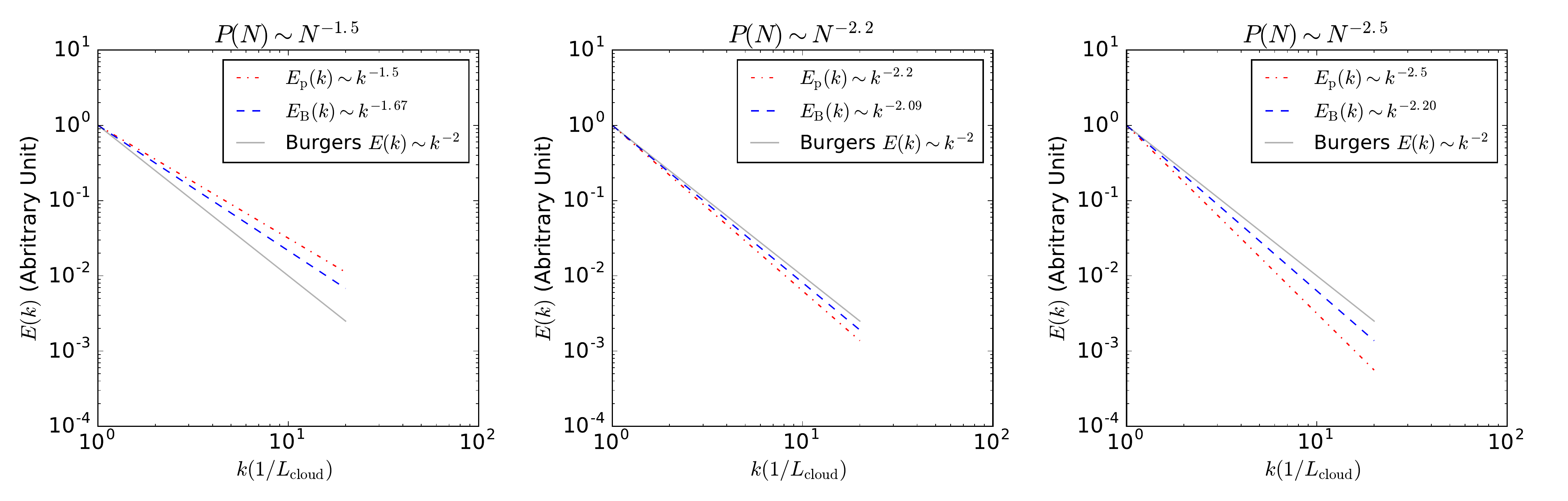}
\caption{Comparison of the gravitational energy spectra and magnetic energy
spectra for different cloud types.
The red lines stand for the gravitational energy spectra and the blue lines
stand for the magnetic energy spectra. The black lines stand for the kinetic
energy spectra of the Burgers turbulence.
The gravitational energy spectra are computed using Eq.\ref{eq:ep},
where different surface density PDFs are assumed. The magnetic energy spectra are computed
using Eq. \ref{eq:eb}, where, additionally, we assume a magnetic field-density
relation $B\sim \rho^{1/2}$.
The left panel shows a g-type cloud where the surface density PDF is
$P(\Sigma)\sim \Sigma^{-1.5}$. The middle panel shows a cloud with $P(\Sigma)\sim \Sigma^{-2.2}$ where the turbulent,
gravitational and magnetic energies are close to each other over multiple scales.
The right panel shows a t-type cloud with a steep surface density PDF
($P(\Sigma)\sim \Sigma^{-3}$). Here we are only comparing the slopes, and the normalisation of the $y$-axis
are not fixed.\label{fig:comp} }
\end{figure*}
\begin{table}
\begin{tabular}{ l l l l }
\hline
$\gamma_{\Sigma}$ & $\gamma_{B}$ & $\epsilon_{\rm p}$ & $\epsilon_{\rm B}$
\\
\hline
\multicolumn{4}{c}{Fiducial case, $B\sim\rho^{1/2}$} \\
  \hline
  1.5  & 0.5 & 1.33  &1.66\ \\
  2 & 0.5 & 2  & 2 \\
  2.5 & 0.5 & 2.4 &  2.2 \\
  \hline
\multicolumn{4}{c}{NGC6334} \\
  \hline
2.26 & 0.4 & 2.23 & 2.5 \\

  \hline
\multicolumn{4}{c}{NGC6334 -corrected, Using Eq. \ref{eq:ep:corr}} \\

  \hline
2.26 & 0.4 & $2.39\pm 0.12$ & $2.5 \pm 0.08$
\end{tabular}
\caption{A list of the slopes derived from observations.
$\gamma_{\Sigma}$ is the slope of the surface density PDF,
$\gamma_{B}$ is the slope of the magnetic field-density relation,
$\epsilon_{\rm p}$  is the slope of the gravitational energy
spectrum and  $\epsilon_{\rm B}$ is the  slope of the magnetic energy
spectrum. The definitions of these quantities can be found in Table
\ref{tbl:def}.\label{tbl:results} }
\end{table}

Assuming that beyond a critical density (typically $n_{\rm H_2}>300\; \rm
cm^{-3}$, \citet{2010ApJ...725..466C}), the magnetic field-density relation $B\sim \rho^{1/2}$
holds universally for all molecular clouds, we discuss the effectiveness of magnetic support in these clouds.
{ A typical molecular cloud should have a surface density PDF of
$P(\Sigma)\sim \Sigma^{-2}$ \citep[e.g.][]{2015A&A...576L...1L}, and this
corresponds to a gravitational energy spectrum of $E_{\rm p}(k)\sim k^{-2}$ (Eq.
\ref{eq:ep}).
Assuming $B\sim \rho^{1/2}$, we can use Eq. \ref{eq:eb} to derive the magnetic
energy spectrum. In this fiducial case, the magnetic energy spectrum is $E_{\rm
B} \sim k^{-2}$. Since both scales as $k^{-2}$, if the gravitational energy and
the magnetic energy of a cloud are comparable on the cloud scale, we expect them
to stay comparable as we move to smaller scales. This is the case where the
gravitational energy and the magnetic energy reach a multi-scale energy
equipartition.
If turbulence in the molecular clouds is Burgers-like, we expect the turbulence
energy to scale as $E_{\rm turb}\sim k^{-2}$, thus there should be a multi-scale
energy equipartition between turbulence, gravity, and the magnetic field.
}

In \citet{2016MNRAS.461.3027L}, molecular clouds are classified into two
categories. The regions whose surface density PDFs are shallower than
$\Sigma^{-2}$ are called g-type clouds. In these regions, the gravitational
energy spectra are shallow, and gravity dominates the cloud evolution on
smaller scales. The regions whose surface density PDFs are steeper than
$\Sigma^{-2}$ are called t-type clouds where turbulence can provide effective
support against gravitational collapse.
The new scaling relations from this paper allow us study the
effect of magnetic field in these different cases. In Fig. \ref{fig:comp}, we
plot the gravitational energy spectrum and magnetic spectrum for different cloud types.
 Assuming that the magnetic field-density relation
$B\sim\rho^{1/2}$ holds universally, both the gravitational energy spectrum and
the magnetic energy spectrum evolves with the cloud surface density PDF.
Therefore, to understand the interplay between gravity and the magnetic
field to a better accuracy, it is necessary to take the variations of the cloud surface
density PDF into consideration.

One should also note that the more recently Bayesian analysis of
\citet{2010ApJ...725..466C} suggest that at densities above  $300 \;\rm
cm^{-3}$, $B\sim\rho^{0.65}$ is the preferred magnetic field-density
relation. What differences does it make when one changes the slope of the
magnetic field-density relation? Consider a typical molecular cloud whose surface density
PDF is characterised by $P(\Sigma)\sim \Sigma^{-2}$, the gravitational energy
spectrum is $E_{\rm p}(k)\sim k^{-2}$ (Eq. \ref{eq:ep}). Assuming
$B\sim\rho^{0.65}$, the magnetic energy spectrum is $E_{\rm B}(k)\sim k^{-1.4}$
(Eq. \ref{eq:eb}). In most cases, the magnetic energy spectrum is much shallower than the gravitational energy spectrum,
which indicates that magnetic fields would play a more important role as one
moves to smaller scales. Since the clouds are already
fragmented, we assume that on smaller ($\sim 0.1\;\rm pc$) scales, the densities
of gravitational and magnetic energy are comparable, on larger scales, we expect the gravitational
energy to exceed the magnetic energy by much. Therefore, a magnetic
field-density relation $B\sim\rho^{0.65}$ would probably imply that magnetic
field is dynamically unimportant on the large scale.
This is similar to the conclusion by \citet{2010ApJ...725..466C}. However, since
the results of \citet{2010ApJ...725..466C}  are obtained by collecting data from clouds with different structures,
it is difficult to evaluate systematic
effects. The current available measurements are consistent with magnetic
fields staying in a multi-scale equipartition with the gravitational energy, but
more future observations should be carried out in a systematic fashion to
further constrain this.
\begin{figure}
\includegraphics[width=0.45\textwidth]{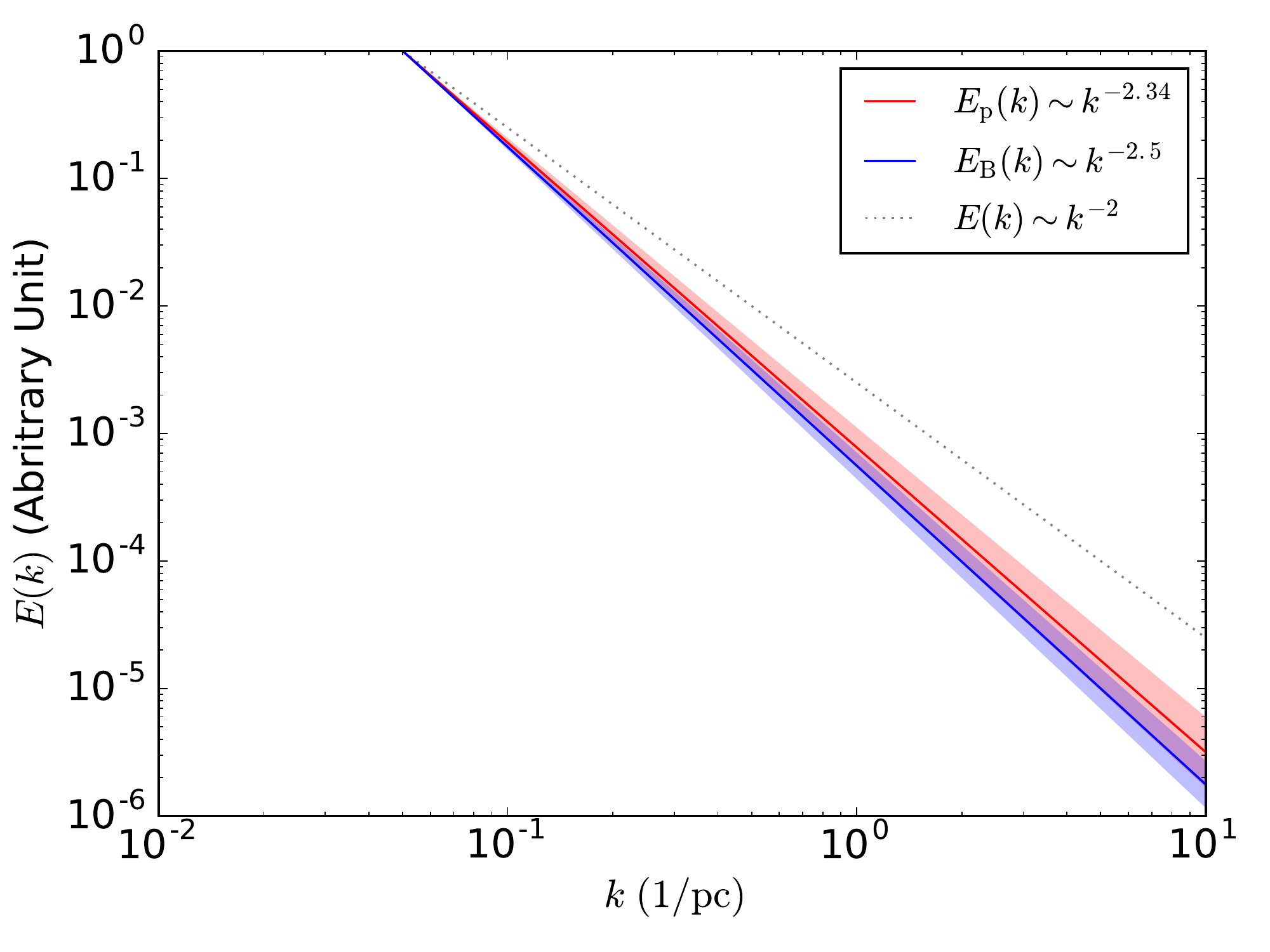}
\caption{\label{fig:ngc6334} The gravitational and magnetic energy spectrum
of the NGC6334 region. The red solid line stands for the gravitational energy
spectrum of the region, the uncertainty of which is represented by the red shaded region.
The blue solid line shows the magnetic energy spectrum of the region, and the
uncertainty is represented by the blue shaded region. The kinetic energy
spectrum of the Burgers turbulence ($E_{\rm k}\sim k^{-2}$) is plotted for
reference.
Here we are only comparing the slopes, and the normalisations of the $y$-axes
are not fixed. See Sec. \ref{sec:ngc6334} for details.
}
\end{figure}
\subsection{ Possible multi-scale energy equipartition in
NGC6334}\label{sec:ngc6334} The NGC6334 star-forming region is perhaps the only region where magnetic
field strength has been {\rm estimated} over multiple scales
\citep{2015Natur.520..518L}, and this allows us to study the multi-scale
interplay between turbulence and gravity to a better detail.
They found that $B \sim \rho^{0.4\pm 0.04}$.
The relation holds from a few tens of a parsec down to $\sim 0.1\;\rm pc$.
However, one
should note that these are not direct measurements: the magnetic field strength
is estimated using force balance, and can be biased. In spite of all these
caveats, the measurements have produced the scaling relation we need to study
the multi-scale interplay between turbulence and gravity.

Since we are aiming at an accurate comparison of the slopes of the gravitational
and magnetic energy spectrum, it is beneficial to take the geometrical
properties of the regions, e.g. how it fragments into account.  The region is fragmented, and
from their Fig.
3, we estimate $N\sim r^{-0.24}\sim k^{0.24}$ where $N$ is the number of regions \footnote{This is derived by
considering the fact that the observations cover 0.1 pc to 10 pc, which
stretch over two orders of magnitude, and the region fragments into three
subregions. $\log_{10}(3) / \log_{10}(100) \approx 0.24$, thus $N \sim
r^{-0.24}$.}.
The surface density PDF of the region has been measured by
  \citet{2013A&A...554A..42R} and  \citet{2015MNRAS.453L..41S} where
\footnote{Here we have used the more recent value presented in \citet{2015MNRAS.453L..41S}.} $P(\Sigma)
\sim \Sigma^{-2.26}$.
The gravitational energy spectrum of the region can be computed with Eq.
\ref{eq:ep}, which is $E_{\rm p}(k) \sim
k^{-2.23}$.
When the number of subregions of the cloud is taken into the analysis, using
Eq.
\ref{eq:ep:corr}, $E_{\rm p, corr}(k) \sim k^{-2.39}$.
 For NGC6334, the
measured magnetic field-density relation is $B\sim \rho^{0.4}$
\citep{2015Natur.520..518L}.
Using Eq. \ref{eq:eb}, it gives a magnetic energy spectrum of $E_{\rm B} \sim
k^{-2.5}$. The gravitational and magnetic energy spectrum are plotted in
Fig. \ref{fig:ngc6334}, and
the scaling exponents are listed in Table
\ref{tbl:results}. The slope of the magnetic energy spectrum
resembles the that of gravitational energy spectrum, suggesting a multi-scale
near equipartition between gravity and the magnetic field in NGC6334.

Although in NGC6334, the slopes still differ by $0.1$, this
difference is not significant had one taken into account the uncertainties: the uncertainty of our formula for the
 slope of the gravitational energy spectrum is found to be
 around $5\%$ for the Orion complex \citep{2017MNRAS.464.4096L}. Since in this
 paper, we take the dependence of the number of regions on the scale
 into account explicitly, we expect a better accuracy. Taking the
 $5\%$ as an upper limit, the gravitational energy spectrum of the NGC6334 region should be
 $E_{\rm p}\sim k^{-2.39 \pm 0.12}$. The slope of the magnetic
 field-density relation measured from \citet{2015Natur.520..518L} has an
 uncertainty of $0.04$. Assuming that this is the major contribution to the
 uncertainty of the magnetic energy spectrum, we have  $E_{\rm B}\sim k^{-2.5 \pm 0.08}$.
 After considering the uncertainties, the results are consistent with an exact
 multi-scale equipartition. between the gravitational and magnetic energy.

\section{Conclusions}
We present an analytical formalism to evaluate the distribution of magnetic
energy over multiple scales in molecular clouds using the observed surface density PDF and the magnetic
field-density relation.

In the fiducial case where a molecular cloud has a surface density PDF of
$P(\Sigma)\sim  \Sigma ^{-2}$ and a magnetic field-density relation of $B \sim
\rho^{1/2}$, the gravitational and magnetic energy reach a multi-scale near
equipartition. For NGC6334 where estimates of
magnetic field over multiple scales (10 pc to 0.1 pc) are available, our detailed analysis
suggest that the energy equipartition
between gravity and the magnetic field holds to a very good accuracy (where the
slopes differ by $\sim0.1$). Thus, if magnetic field is
dynamically important on the cloud scale, we expect it to be important on much
smaller scales, and would be able to influence the gravitational collapse over
these scales.


Our analysis also reveal our lack of quantitative understanding of the role of
magnetic fields in clouds: The slope of the magnetic field-density relation
remains uncertain. Besides, according to our formalism, to properly evaluate
the gravitational and magnetic energy densities as a function of the scale, one
needs to measure both the surface density PDF and the magnetic
field-density relation within different clouds. This should be achievable in the
future.

It is seems likely that many more clouds would have energy
equipartition similar to the NGC6334. However, if future observations indicate
a steep slope in the magnetic field-density relation -- a slope that is steeper than what is predicted by Eq.
\ref{eq:equi}, it implies that there is an excess of magnetic energy
density on smaller scales compared to the cloud scale, and presumably magnetic
field is not important on the large scale, but becomes important as it gets
amplified by some dynamo processes \citep[e.g.][]{1993IAUS..157..429L} during
the collapse.
On the other hand, if observations indicate a shallower slope,
it would imply that the gas contained in small-scale structures has lost a significant fraction of the magnetic
flux during the collapse, and in this case the magnetic field is
probably regulating the collapse. The importance of the magnetic field might
also be different in different clouds. Future
observations should be able to to distinguish these possibilities.\\

\noindent
{\it
Notes after the proof:}\\
\noindent
(1) Although the concept of magnetic energy spectrum is developed based on our earlier papers, a similar construction has been proposed by \citet{2004ApJ...610..820H} (reviewed by \citet{2017ARA&A..55..111H}), where, using pulsar rotation and dispersion measures, they found $E_B(k) \sim k^{-5/3}$. Their results are obtained using a different method, yet the values are consistent with ours.\\
\noindent
(2) The equipartition of the spectral energy densities of the turbulence and gravity has been found in the simulation of \citet{2017NJPh...19f5003K}.

\section*{Acknowledgements}
 Guang-Xing Li is supported by the Deutsche
Forschungsgemeinschaft (DFG) priority program 1573 ISM- SPP.
Guang-Xing Li would
like to thank the referee of \citet{2016MNRAS.461.3027L} for suggesting the
analysis of the magnetic fields, and would like to thank our referee for
very helpful reports and suggestions.








\appendix

\bibliography{paper}


\bsp	
\label{lastpage}
\end{document}